\documentclass[  journal=largetwo,  manuscript=article-type,  year=2020,  volume=37,]{cup-journal}

\usepackage{amsmath,amssymb}
\usepackage[nopatch]{microtype}
\usepackage{booktabs}
\usepackage{longtable}
\usepackage{tablefootnote}

\newcommand{\kms}{km s$^{-1}$}
\newcommand{\ms}{ms$^{-1}$}
\newcommand{\ps}{s$^{-1}$}

\title{Evolution of the magnetic field and flows of solar active regions with persistent magnetic bipoles before emergence.}

\author{C.S.~Alley}
\affiliation{School of Information and Physical Sciences, The University of Newcastle, Australia}
\author{H.~Schunker}
\affiliation{School of Information and Physical Sciences, The University of Newcastle, Australia}
\email[H. Schunker]{hannah.schunker@newcastle.edu.au}
\newcommand{\aap}{{Astron. Astrophys.}}

\newcommand{\apjs}{{Astrophys. J. Supp.}}
\newcommand{\apjl}{{Astrophys. J. Lett.}}

  %% First letter not capped

\begin{document}

\begin{abstract}
Magnetic active regions on the Sun are harbingers of space weather. Understanding the physics of how they form and evolve will improve space weather forecasting. 
Our aim is to characterise the surface magnetic field and flows for a sample of active regions with persistent magnetic bipoles prior to emergence.
We identified 42 emerging active regions (EARs), in the Solar Dynamics Observatory Helioseismic Emerging Active Region  survey (Schunker \textit{et al.} 2016), associated with small magnetic bipoles at least one day before the time of emergence. We then identified a contrasting sample of 42 EARs that emerge more abruptly without bipoles before emergence. We computed the supergranulation scale surface flows using helioseismic holography. We averaged the flow maps and magnetic field maps over all active regions in each sample at each time interval from 2~days before emergence to 1~day after.
We found that EARs associated with a persistent pre-emergence bipole evolve to be, on average, lower flux active regions than EARs that emerge more abruptly.
%$(19\pm1.3)~\times 10^{20}$~Mx, compared to, $(24\pm1.8) \times 10^{20}$~Mx. 
Further, we found that the EARs that emerge more abruptly do so with a diverging flow of $(3\pm 0.6) \times 10^{-6}$~\ps on the order of 50-100~\ms.
Our results suggest that there is a statistical dependence of the surface flow signature  throughout the emergence process on the maximum magnetic flux of the active region.
\end{abstract}
% 220 words
% HS: check outputs in code plot_ave_flux_flows_LBP.pro for the flow values

\section{Introduction}\label{sect:intro}
 Active regions are generally thought to be formed by coherent, arched magnetic flux tubes rising through the interior to manifest as roughly east-west aligned opposite polarity pairs at the surface of the Sun. In many dynamo models they are important to convert the Sun's global toroidal magnetic field to poloidal \autocite{CS2015}. Understanding  the physics behind the emergence process is important to constrain their origins and connection to the Sun's large scale global field, as well as for space weather forecasting.

It is not clear from what depth these flux tubes originate, nor
what causes them to rise. At the surface and below, the Sun’s
magnetic field is embedded in the convective flows, and as such
can be modelled with the equations of magnetohydrodynamics.
Below the surface, the plasma pressure is generally greater than
the magnetic pressure. Understanding the dominant terms in
these equations, particularly the flows and the magnetic field, is important to understand how active regions form.

Simulations of magnetic flux emergence show that thin flux tubes can rise a priori from the base of the convection zone due to magnetic buoyancy in the absence of convection \autocite[e.g.][]{Fanetal1993,Weber2011}; or tubes of magnetic flux can be formed within the convection zone itself and brought up by convective upflows  \autocite[e.g.][]{Chenetal2017,HottaIijima2020}. It may also be possible for active regions to form without the presence of a flux tube \autocite[e.g.][]{Brandenburg2005,Brandenburgetal2014}. 
%The generation of the flux tubes prior to their emergence is another question. 
For a full review see \cite{FanLRSP2021}.

The thin flux tube models in \cite{Fan2008} predict a retrograde flow at the peak of the flux tube just before emergence. In principle, local helioseismology  could be used to measure this flow. 
Local helioseismology measures perturbations to the natural seismic oscillations driven by the turbulent convection at the surface of the Sun and infers the three-dimensional subsurface structure and dynamics in localised areas \autocite[see ][ for a full overview]{GizonBirch2005}.
\cite{Birchetal2013} used helioseismic holography \autocite{LindseyBraun2000} to measure the subsurface flows prior to the formation of one hundred active regions observed by the Global Oscillation Network Group \autocite[GONG;][]{Harveyetal1998}. They found that there were no statistically significant flows below the surface, however near the surface they found a statistically significant flow of about 15\ms towards the emergence location in the day preceding the active region formation. 

Recently, it has become apparent that the near-surface convective flows themselves are important in the emergence process \autocite[see][for a summary of the recent paradigm shift]{Weberetal2023}. By comparing the observed surface flows at the time of active region emergence with simulations, \cite{Birchetal2016} showed that flux tubes cannot be rising faster than about 100~\ms through the upper convection zone, on the order of the convective velocities themselves.  \cite{Birchetal2019} went on to show that, on average, active regions preferentially emerge in east-west aligned converging flow lanes, suggesting a connection to the Sun's supergranulation pattern. \cite{Schunkeretal2019} showed that the growth of active regions through the emergence process is consistent with the length and timescales of supergranulation, supporting the idea that convection on these spatial scales buffets the polarities as they emerge.

From a practical perspective, perhaps most importantly for space weather, it would be useful to be able to predict the location and time an active region will emerge, and whether or not it is likely to have low magnetic flux or high magnetic flux. Some effort has been dedicated to this. Helioseismic studies have not resulted in a significant subsurface detection \autocite[see e.g.][]{Kommetal2015}, but \cite{Barnesetal2014} showed that in the day before emergence the strongest indication that an active region will emerge is the unsigned surface magnetic field itself. This was also reflected in more recent efforts using machine learning \autocite{Dhurietal2020}. 
%While this could suggest that a better definition of `emergence' is required, these active regions  deserve a more detailed understanding. 

In this paper, we examine the magnetic field prior to emergence for two samples of active regions. In Section~\ref{sect:hears} we describe our database of  emerging active regions. We then outline the data reduction, computation of the surface flows, how we measure the location of the polarities in each active region, and the ensemble averaging of the flow and magnetic field maps (Section~\ref{sect:datared}). In Section~\ref{sect:bipoles} we describe the identification of active regions that show persistent pre-emergence bipole magnetic fields. In Section~\ref{sect:aves} we show that there is a distinct difference in the averaged surface flows and magnetic field from $2$~days before the emergence to 0.6~days afterwards, and in Section~\ref{sect:disc} we discuss the implications of our results and possible avenues to make future progress.

\section{Sample of observed emerging active regions}\label{sect:hears}
% hears
The Solar Dynamics Observatory Helioseismic Emerging Active Region (SDO/HEAR) Survey currently consists of 182 emerging active regions  (EARS) \autocite[for a full list see][]{Schunkeretal2016, Schunkeretal2019} observed by the Helioseismic and Magnetic Imager onboard the Solar Dynamics Observatory \autocite[\textit{SDO/HMI}, ][]{Scherreretal2012}  between May 2010 (the start of science grade SDO/HMI observations) and July 2014 (the declining phase of solar cycle 24). To avoid complications in interpreting local helioseismology results \autocite[e.g.][]{Schunker2010}, the active regions selected for the HEAR survey specifically emerge into a relatively quiet area of the Sun’s surface at least $20^\circ$ from any surrounding strong surface magnetic fields.
%The HEAR survey only includes active regions with National Oceanic and Atmospheric Administration (NOAA) numbers that reached a total area (in continuum intensity) of at least 10~$\mu$H (micro hemispheres, 1~$\mu$H $\approx 3~$Mm$^2$).  

%define emergence time
The emergence time for each active region, $t=0$~days (see \ref{app:lbp} and \ref{app:nlbp}), is defined as the time when the absolute flux, corrected for line-of-sight projection, reaches 10\% of its maximum value over a 36~hour interval following the first appearance of the sunspot (or group) in the National Oceanic and Atmospheric Administration (NOAA) record \autocite{Schunkeretal2016}. We emphasise that a small amount of flux from the EAR is present at the emergence time. A negative time indicates pre-emergence and a positive time indicates post-emergence for each EAR.

%naming convention
Each active region, labelled by its NOAA number,  is paired with a corresponding control region (CR). The CR is assigned a mock-emergence time when the Stonyhurst coordinates of the CR was the same as the EAR at its (real) emergence time \autocite[see Table~\ref{tab:lbp} and Table~\ref{tab:nlbp} in this paper, and ][ for a full description]{Schunkeretal2016}. The control region is necessary to ensure that the signal is not due to systematics from the SDO/HMI instrument or due to some line-of-sight observing effect.

\section{Data reduction}\label{sect:datared}
Our data reduction closely follows what has already been published in \cite{Schunkeretal2016}, \cite{Birchetal2016} and \cite{Birchetal2019}. Here we outline the relevant details for this study.

\subsection{Mapping and tracking}\label{subsect:maptrack}
The HMI observes the full-disk continuum intensity, line-of-sight magnetic field and Doppler velocity at the photosphere with a cadence of 45~s. 
% tracking
We track the location of the EARs and CRs at the Carrington rotation rate over time intervals (\texttt{TI}) of 6.825~h (547 frames with a cadence of 45~s).  

Local helioseismology measures the travel time of a wave from one location on the surface until it appears at another location on the surface. Waves travel at the local sound speed, and generally waves that travel shorter distances do so over a shorter time. The waves we are interested in for this study have travel times less than about 6~hours, and so this is the length of time we chose for a single time interval and the corresponding datacube.

These datacubes are labelled with their time interval (\texttt{TI}) relative to the emergence time interval, \texttt{TI+00}. The emergence time is defined as $\tau=0.0$~days, and the midtime of time interval \texttt{TI+00} corresponds to $\tau=0.1$~days \autocite[see Table~B.1 in ][ which lists the mid-time of  each \texttt{TI} relative to the time of emergence, $\tau=0$]{Schunkeretal2019}.  
The beginning of each time interval is spaced at 5.3375~h (320.25 min, 427 frames), and they have a 1.5~hour, or 120 frame, overlap \autocite[see Figure~5 in ][]{Schunkeretal2016}.
Each active region is tracked up to 7~days before and after the emergence time, depending on their distance to central meridian at that time.

% mapping
At each 45~s interval we projected the full-disk SDO/HMI observations onto  $60^\circ \times 60^\circ$ Postel projection maps. The projection is made to a $512 \times 512$~pixel grid with a pixel size of 1.39~Mm. The coordinates of the map centre are the flux-weighted centre of the line-of-sight magnetic field at the emergence time \autocite[see ][ for more details]{Schunkeretal2016}.
In this article we examine the magnetic field maps averaged over each 6.825~hour time interval to correspond directly to the flows.

\subsection{Computing the surface flows}
%how to get flows from doppler
Local helioseismology is a tool that uses the acoustic waves in the Sun to map the three-dimensional subsurface structure and/or dynamics (for a review of the different methods and key results see \cite{GizonBirch2005}). For example, perturbations to the travel-times of the waves can be interpreted as a linear perturbation to the structure and dynamics of the interior of the Sun at the depths where that particular wave has sensitivity.
By selecting waves that are sensitive to the  near-surface of the Sun, we can infer the horizontal flows which we can then be used to identify supergranulation structures \citep[e.g.][]{Gizonetal2003}.

We filtered the tracked and remapped Doppler velocities with a phase-speed filter with a central phase speed of $17.49$~\kms and a width of $2.63$~\kms \autocite[filter 3 from Table~1 in ][]{Couvidatetal2005}. This filter isolates waves that are most sensitive to the 3~Mm just below the photosphere. We then measured the north-south and east-west travel time differences using surface-focusing helioseismic holography \autocite{LindseyBraun2000}. We used an empirically determined conversion constant of $-7.7$~ms$^{-2}$ to convert from travel time differences to surface flows \autocite{Birchetal2016}.
% AARON USES A CONVERSION CONSTANT OF -5.7 ms-2 empirically determined from LCT flows

%we multiply the vx and vy by a factor for TD3 REFERENCE PAPER -7.69553\\
%We convert the oi travel time to divergence by dividing by pixel size\\
%we remove a fit to the 2d background exlcuding the cntral 50 pixels\\
%we filter the high wavenumber components using a filter which is 1 for $k R_\odot < 140$ and goes to zero via a raised cosine at $k R_\odot = 220$\\

We remove any remnant large scale velocities from the Sun's bulk rotation or orbital velocity of the SDO satellite (a constant offset) in the flows by subtracting a plane fit to each map, excluding a central region of radius 70~Mm (50~pixels) which is our region of interest where the active regions emerge. 
To remove high-wavenumber noise, we use a low-bandpass filter with a value of 1 for $k R_\odot < 140$, a raised cosine from 1 to 0 in the region $140 < k R_\odot < 220$, and zero for $k R_\odot > 220$. 
We then have surface flow maps, $\mathbf{v_x}(x,y)$ and $\mathbf{v_y}(x,y)$ at a 5.3375~h time interval, corresponding to the time- averaged magnetic field maps (described at end of Section~\ref{subsect:maptrack}).
%We tested that our results were consistent using  flows computed using local correlation tracking \autocite{NovemberSimon1988,Birchetal2016}.

%\subsection{Ensemble averages}
%summarise data
To create ensemble averages, we treat all active regions as if they were in the northern hemisphere, so that positive $\mathbf{y}$, is towards the pole (north);  negative $\mathbf{y}$ is towards the equator (south); positive $\mathbf{x}$ is in the prograde (solar west) direction; and negative $\mathbf{x}$ is in the retrograde (solar east) direction.

%   correction to flow and B maps
We reversed the magnetic field polarity of the regions in the southern hemisphere, to account for Hale’s law when averaging EARs, so that the leading polarity is always negative. Under the assumption that the magnetic field is radial at the solar surface, we approximately corrected for the magnitude of the magnetic field for the line-of-sight projection by dividing it by $\cos \theta$, where $\theta$ is the angular distance to disk centre. 

For active regions in the southern hemisphere, we flipped the averaged magnetic field and flow maps in the latitudinal direction to account for the pole-to-equator symmetry, and reversed the direction of the flows in the north-south direction, so that the poleward flows are in the positive $y$-direction. We then computed the divergence of the flows, $\bf{\nabla} \cdot \bf{v}_h$ where $\bf{v}_h =(v_x, v_y)$, as the most representative way to show the location of the supergranules, which are the tops of convection cells. 

\subsection{Measuring the location of the bipoles}\label{sect:pos}
The magnetic field associated with an emergence can have significant proper motion compared to the Carrington rotation rate. To analyse the evolution of the flows associated with the EARs, we measured the location of the active region magnetic field at each time interval.

% centroid of each polarity from 2019 paper
We tracked the position of the centroid of the positive and negative polarity in the active region as described in \citet{Schunkeretal2019}, and we outline the process here. We first measured the location of the roughly circular polarities with a threshold magnetic field strength of 20~G  at time interval \texttt{TI+02} ($\tau=0.6$~days). We used a feature recognition algorithm (feature.pro copyright 1997, John C. Crocker and David G. Grier)  designed to determine the centroid position of roughly circular features in an image to determine the location of both polarities individually. Moving forward and backwards in time, we repeated the process and selected the $x$ and $y$-centroid closest to the polarity location in the previous time interval.

If the location of the bipole at some time interval is not defined (e.g. a bipole cannot be detected), then we linearly interpolated for the $x$ and $y$-centroid locations from the nearest time intervals. For times before a clear bipole is detected, we extrapolate the first measured location of the bipole. Similarly, for times after emergence when a clear bipole can no longer be detected, e.g. after decay, we extrapolate the last measured location of the bipole. 

We shifted the averaged magnetic field and flow maps, using a bi-linear interpolation (over the nearest four pixels), for each EAR to the point half-way between the centroid locations at each time interval.  
The shifts are typically on the order of up to $5$~pixels. The shifting of the maps at each time interval removing the proper motion of the bipole is unique to the analysis method in this paper, and is required to get a well-defined mean of the absolute magnetic field prior to emergence.

\begin{figure*}
	\includegraphics[width=1.0\textwidth]{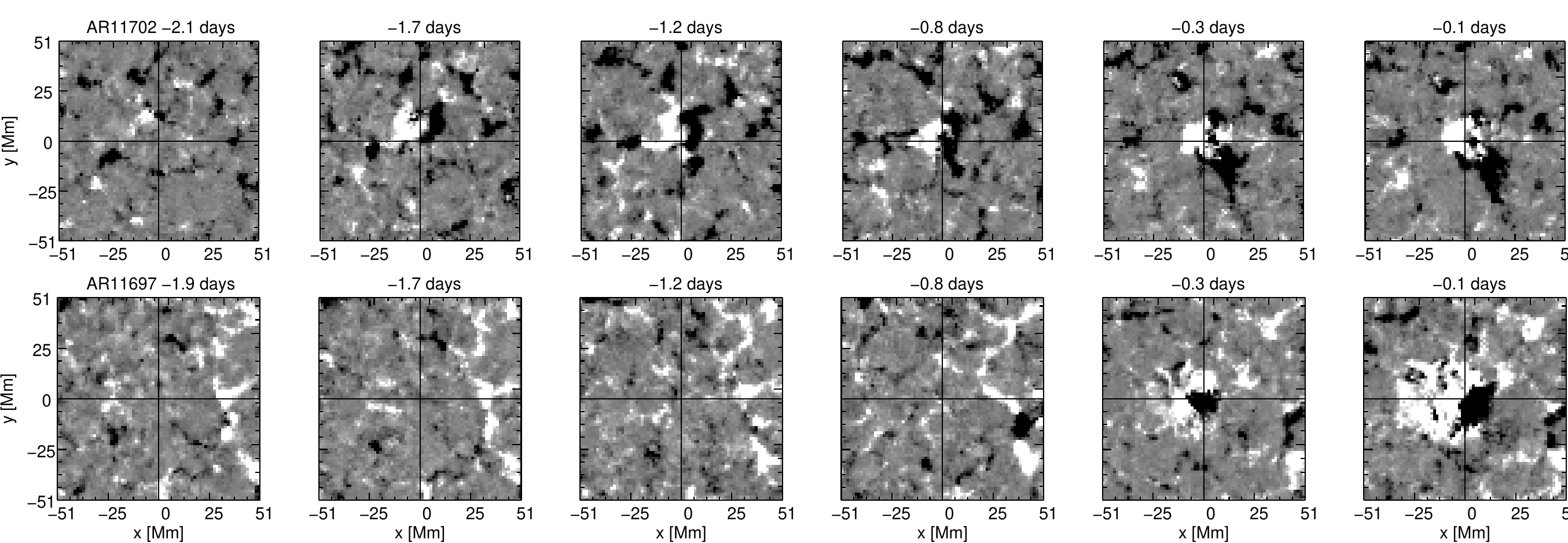}
    \caption{Time averaged line-of-sight magnetic field maps for an
    example EAR with a bipole associated with the emerging flux more than two days before the emergence time (top row, AR~11702). And an example of an EAR without an obviously associate pre-emergence bipole  (bottom row, AR~11697). The greyscale is saturated at $\pm 15$~G. Note that these maps have not been shifted to the emergence location.}
    \label{fig:examplelbp}
    % figure_AR11702_AR11697_bmaps.pro --run at MPS, tyring to get fitsio_read_image.pro to work in laptop
\end{figure*}

\section{Identification of persistent magnetic bipoles before emergence.}\label{sect:bipoles}

Active regions are defined by their dark appearance in the intensity of light from the surface of the Sun. This is due to the strong concentrations of magnetic field, that appear in an east-west aligned pair of opposite polarities on the solar surface, and increases in size and field strength as more flux emerges \autocite[see, for example, Figure~1 in][]{Schunkeretal2019}. It takes about 2~days on average from the time of emergence for an active region to reach its maximum flux \autocite[see Figure~2 in][]{Weberetal2023}.

% see AR Catalog with Comments.xls in the shared drive "Camron" directory
By inspection, we identified 42 EARs (listed in ~\ref{app:lbp}) as having persistent magnetic bipoles at least one day before emergence associated with the eventual bipole structure of the active region. The persistent polarities are characterised by their roughly east-west orientation and the proximity as a pair as they are buffeted by the convection. The bipoles do not change significantly in size (on the order of 10~Mm, see Figure~\ref{fig:examplelbp}, first panel) until the main emergence process begins closer to $t=0$. 

Some active regions, such as AR~11182 and AR~11969, show magnetic bipoles up to 2.8~days (\texttt{TI-12})  before emergence. An example of an EAR with a clearly associated pre-emergence bipole more than 2~days before emergence is AR~11702 shown in the top row of Figure~\ref{fig:examplelbp}.

For a contrasting sample, we selected an equal number (42) of EARs that, by inspection, are not associated with any magnetic field bipoles prior to $t=-0.3$~days (listed in ~\ref{app:nlbp}). One example is AR~11697 shown in the bottom row of Figure~\ref{fig:examplelbp}. These two samples may constitute the extremes of a continuum  allowing us to clearly identify any fundamental differences in their evolution.

Any EARs with  dense, small-scale magnetic field within $\approx50$~Mm radius of the emergence location were excluded from either sample since they may obscure the identification of any pre-emergence bipole.
% maybe a referee will ask for an example of one that was excluded?
In \cite{Schunkeretal2016}, the authors defined a $P$-factor, where 0 represents an emergence into a very quiet region; a $P$-factor of 1 or 2 indicates emergence into increasing amounts of magnetic field nearby (but not directly at) the subsequent emergence location; and a $P$-factor of 3 or higher indicates the region may be compromised by pre-existing field at the emergence time and location. 
We cross-checked the $P$-factor for the active regions identified in our two samples, and found that  neither of the samples we identified have a dominant $P$-factor, and were mostly 0 or 1.
 
In summary, we have averaged line-of-sight magnetic field maps, $B(x,y)$, flow maps, $v_x(x,y)$ and $v_y(x,y)$, and flow divergence maps $\bf{\nabla} \cdot \bf{v}_h$ with a 5.3375~h cadence, centred on the bipole location at each time interval for 42~EARs with persistent pre-emergence bipoles, and 42~EARs without pre-emergence bipoles.

\begin{figure*}
	\includegraphics[width=0.99\textwidth]{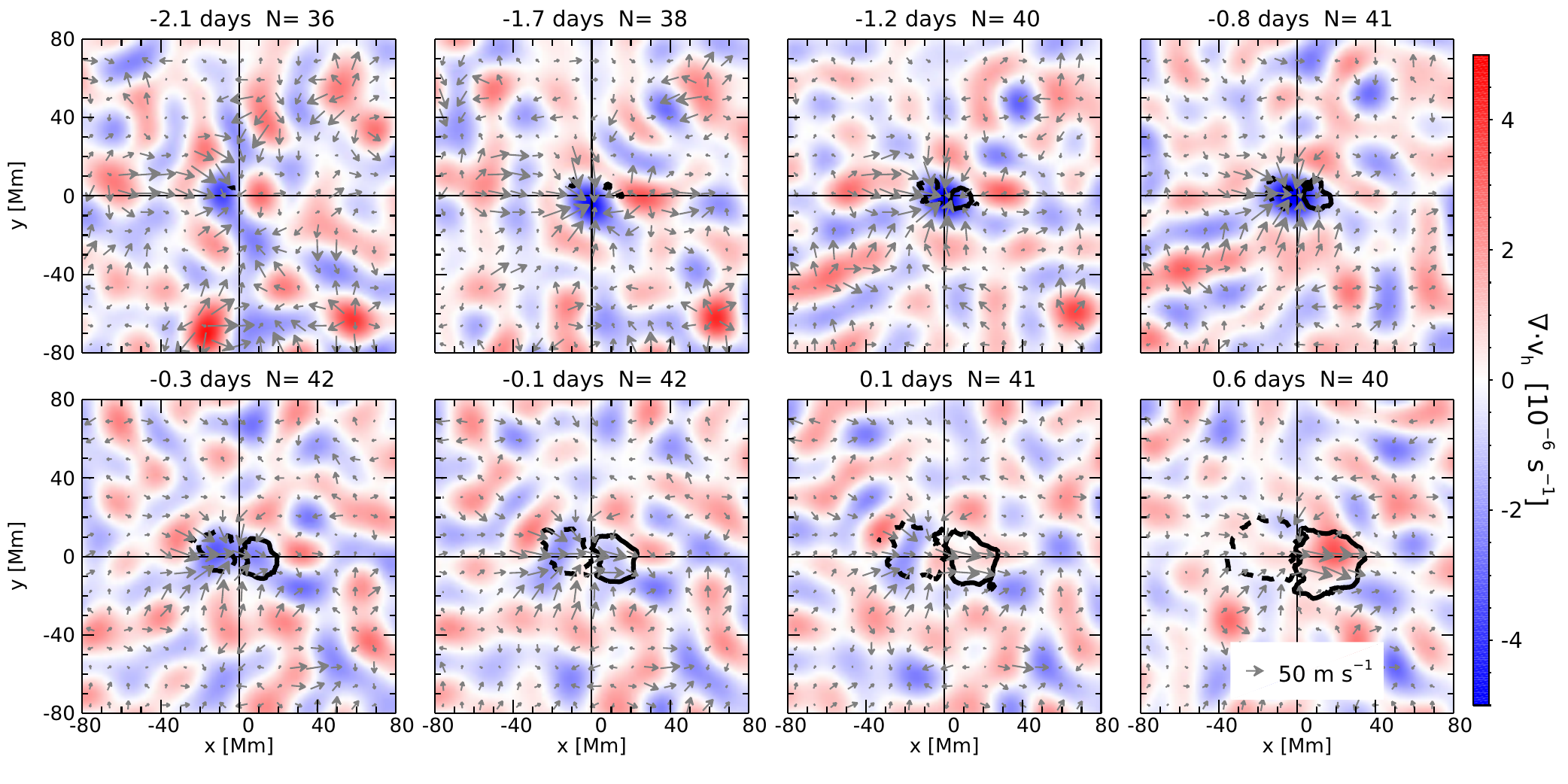}
    \caption{Averaged divergence flow maps over $N$ active regions with pre-emergence bipoles at different time intervals. Blue represents converging flows and red represents diverging flows. The arrows indicate the direction and magnitude of the flows. Solid (dashed) black lines contour  the $-20$~G ($+20$~G) of the averaged line-of-sight magnetic field maps. There is a significant converging flow prior to emergence. }
    \label{fig:avelbpmaps}
\end{figure*}
% ave_maps_EARS_lists.pro
% image_ave_EARS_lists.pro
% figure_ave_EAR_lists.pro
\begin{figure*}
    \vspace{1.0cm}
	\includegraphics[width=0.99\textwidth]{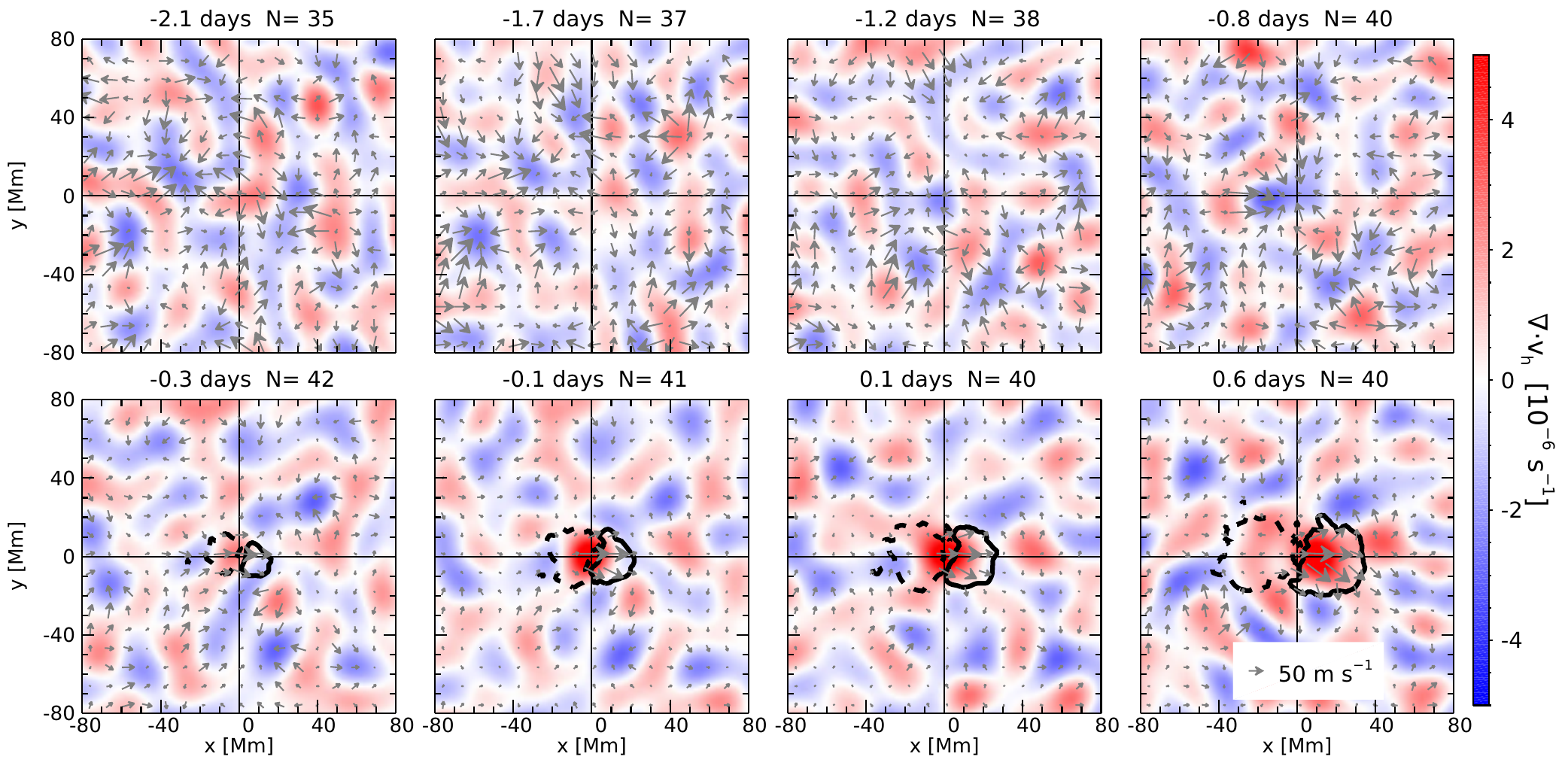}
    \caption{Averaged divergence flow maps over $N$ active regions without pre-emergence bipoles at different time intervals. Blue represents converging flows and red represents diverging flows. The arrows indicate the direction and magnitude of the flows. Solid (dashed) black lines contour $-20$~G ($+20$~G) of the averaged line-of-sight magnetic field maps. There is no significant flow signal prior to emergence, however, there is a significant diverging flow post-emergence. }
    \label{fig:avenlbpmaps}
\end{figure*}
% ave_maps_EARS_lists.pro
% image_ave_EARS_lists.pro
% figure_ave_EAR_lists.pro
%\clearpage 

\section{Evolution of the average magnetic field and flows}\label{sect:aves}
We averaged the magnetic field maps and the flows for each sample of EARs (with and without persistent pre-emergence bipoles). Figures~\ref{fig:avelbpmaps} and \ref{fig:avenlbpmaps} show the average surface flows at each time interval for the two samples. 
The equivalent maps for the control regions are in Figures~\ref{fig:avelbpqsmaps} and ~\ref{fig:avenlbpqsmaps} in \ref{app:aveqs}.

The sample of EARs with persistent pre-emergence bipoles (Figure~\ref{fig:avelbpmaps}) shows a converging flow at the location of the bipoles (centre of the map) prior to the emergence time. The growth of the bipole is shown in the contours of the mean magnetic field.
The sample without pre-emergence bipoles (Figure~\ref{fig:avenlbpmaps}) does not show a statistically significant converging flow, but does show a statistically significant diverging flow from the time of emergence. 

% surface B spatially averaged at the centre
At each time interval we computed the spatially averaged magnetic flux within a central disk of radius 35~Mm for each active region. We chose a fixed radius that encompasses the contour of the average $|B_\mathrm{los}|=20$~G as the active region grows in size to $\tau=0.6$~days (see Figure~\ref{fig:avelbpmaps}).

\begin{figure*}
	\includegraphics[width=1\textwidth]{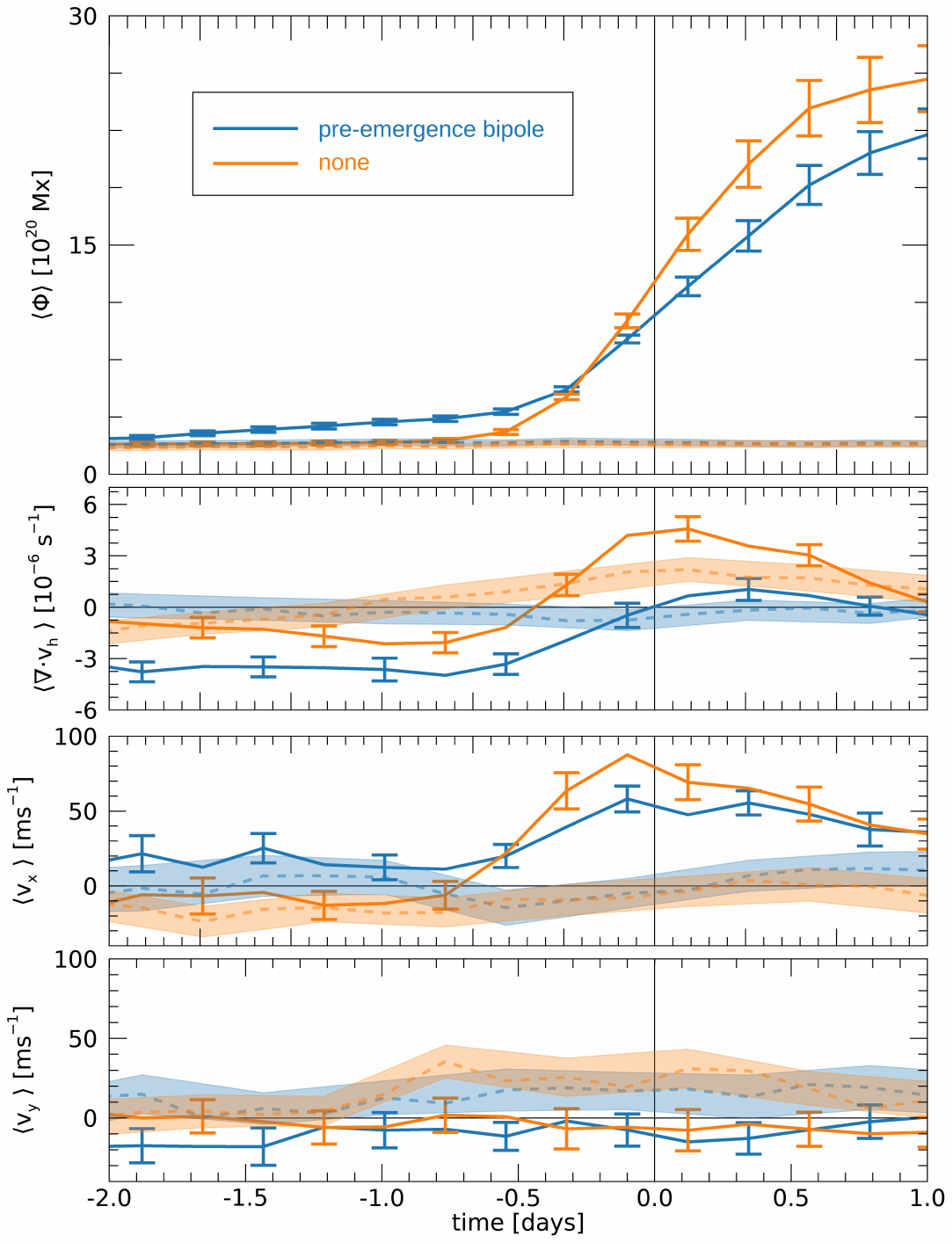}\vspace{-1.5cm}
    \caption{ Averaged magnetic flux and flows as a function of time for active regions with pre-emergence bipoles (blue) and those without (orange). 
    The top panel shows the average magnetic flux within a central 35~Mm radius of the emergence location as a function of time. The error bars show the standard error in the average magnetic flux over the active regions at each time interval. The dashed lines show the corresponding control regions with shaded  standard errors. 
    Active regions with pre-emergence bipoles tend to evolve to be lower magnetic flux active regions post-emergence.
    The second panel shows the averaged flow divergence, the third panel shows the average East-West flow, and the bottom panel shows the average North-South flow. 
    The surface flows are averaged in the central 11~Mm radius of the emergence location.
    }
    \label{fig:aveflows}
\end{figure*}

The top panel of Figure~\ref{fig:aveflows} shows the mean and standard error of the magnetic flux over the sample of active regions with (blue), and without (orange), a pre-emergence bipole. The expected difference in magnetic flux prior to the emergence of the active regions is clear. 
Then, after about -0.3 days, the averaged magnetic flux of the regions that emerge abruptly becomes higher than for the regions with pre-emergence bipoles. This shows that the active regions with pre-emergence bipoles evolve to have significantly lower flux than those without pre-emergence bipoles.

% surface flows spatially averaged at the centre
We then averaged the surface flows in the central 11~Mm (8 pixels) radius for each active region and time interval. We chose this radius by inspection of the maps in Figure~\ref{fig:avenlbpmaps} to include only the flows associated with the emergence, and note that it is on the order of supergranulation scales, but is considerably smaller than the radius of the area over which the magnetic flux was averaged (35~Mm). The lower three panels of Figure~\ref{fig:aveflows} 
show the evolution of the averaged flow divergence, east-west velocity, and north-south velocity. The averaged flow divergence follows a similar evolution for both populations apart from an offset of about $3 \times 10^{-6}$\ps.
The mean flow in both samples is converging up until 0.5~days before emergence, $\langle \mathbf{\nabla} \cdot \mathbf{v_h} \rangle = (-2.9  \pm 0.5) \times 10^{-6}$~\ps  for regions with a pre-emerging bipole, and $\langle \mathbf{\nabla} \cdot \mathbf{v_h} \rangle = (0.03  \pm 0.4) \times 10^{-6}$~\ps for those without. 
From about 0.5~days before emergence until about 0.1 days after emergence, the divergence increases in both samples. 

The flows for both samples are strongest in the east-west direction, $\langle \mathbf{v_x} \rangle$, and peak close to the time of emergence at about $70$~\ms. The flows in the north-south direction, $\langle \mathbf{v_y} \rangle$, are consistent with zero and do not vary significantly, due to averaging over the antisymmetric flow in the north-south direction (see Figure~\ref{fig:avelbpmaps}).
% (the difference is $\langle \Delta v_\textrm{oi} \rangle =3.0  \pm 0.5 \times 10^{-6}$~/s).
% see print to screen from plot_ave_flux_flows_LBP.pro

 Most of the surface of the Sun consists of the diverging flows of granules and supergranules.  Statistically, it is more likely that a randomly selected location on the surface will be a diverging flow, rather than the narrower inflow lanes. This is reflected in Figures~\ref{fig:avelbpqsmaps} and ~\ref{fig:avenlbpqsmaps} (which shows the averaged surface flow maps of the control regions), and the orange dashed curve in the second panel of Figure~\ref{fig:aveflows}. The averaged divergence signal close to the emergence time in Figure~\ref{fig:aveflows} is equivalent in magnitude to other diverging flows in the map (see ~\ref{app:aveqs}), showing that it is not significant compared to the background signal. Because we  averaged the flows over a relatively small sample of active regions and the averaging area (diameter of about 20~Mm) is the size of a supergranule, we have captured the evolution of supergranulation-scale flows.

% ave_maps_EARS_lists.pro
% image_ave_EAR_lists.pro
% plot_ave_flux_flows_LBP.pro
% plot_ave_flux_flows_LBP.eps

\section{Discussion}\label{sect:disc}

From our classification of two samples of EARs based on their pre-emergence bipole signatures we found that these samples also form distinct post-emergence populations of active regions. The sample of active regions with persistent pre-emergence bipoles evolve to be, on average,  lower flux active regions\footnote{We do note, however, that although the largest active region in our sample, AR~11158, falls in this category, it is a double emergence where two bipoles emerge close to one another and then recombine to form a large, complex active region. In this study, we have effectively only followed the central bipole.}, and those that emerge more abruptly are higher flux active regions. It is not clear why this is the case, but suggests that there is some flux-dependence in the growth and evolution of active regions.

Our results are consistent with \cite{Birchetal2019} who show that active regions tend to emerge into regions of converging flow, but we add that the amplitude and sign of the flow divergence is flux dependent. We find that the sample with 
persistent pre-emergence bipoles emerges into strong converging flows and these flows begin more thsn two days prior to emergence. Pre-emergence bipoles  confined to the converging flow lanes between supergranules is not surprising since  small, lower flux magnetic features are buffeted by the flows at the surface of the Sun.

Secondly,  \cite{Birchetal2016} showed that the upward rise speed of flux tubes that form active regions must be on the order of the convective flow velocities ($\lesssim 100$~\ms), based on the lack of any diverging flow signature in a single observed active region. In our statistical analysis, we have shown that some samples of active regions \emph{are} associated with a diverging flow, although the magnitude  is on the order of supergranulation velocities ($\approx 100$~\ms) and not as strong as the diverging flows produced by simulations of a traditional thin flux tube  \autocite[up to $500$~\ms][]{Birchetal2016} .

Furthermore, this sample of EARs shows that higher flux active regions emerge with stronger diverging flows. This may suggest that it is easier to form a large active region where the converging (down) flow is weak, and/or that the magnetic flux is brought up in the upflows of newly forming supergranules. On the other hand, it could also suggest that  tubes with higher magnetic flux rise faster, driving a moderate surface diverging flow at the time of emergence, however this can only be properly explored once the emergence process of the full sample of active regions is (better) understood.

We suspect that these active regions with pre-emergence bipoles are responsible for the conclusions of \cite{Barnesetal2014} and \cite{Dhurietal2020} that the best predictor of an imminent active region emergence is the surface magnetic field itself. 
It may also be that these active regions are at the sites of active region `nests' \autocite{Isik+20} or active longitudes \autocite[e.g.][]{BerdUso2003} supplementing the surface small-scale magnetic field from below. 
A larger sample of active regions will help to explore this idea.  
%May be associated with active longitudes. Haven't checked.

We also note that we are limited by our resolution of 1.39~Mm per pixel, which is four times coarser than the nominal HMI resolution. From inspection of some of the full resolution line-of-sight magnetograms for these active regions, the onset of the bipoles before emergence and their motions can be tracked more precisely. 

\cite{HottaIijima2020} have successfully formed a small active region by placing a flux tube at some depth in their three dimensional numerical magnetohydrodynamic simulations where the near-surface convective flows have brought the flux to the surface. The key to the formation of their active region is placement in an upflow region sandwiched between two downflow regions. The emergence in an upflow region is consistent with our sample of EARs with higher flux, but without pre-emergence bipoles. These simulations are computationally expensive, and while an equivalent statistical sample of emerging active regions to compare with is prohibitive, such simulations are crucial to probe the subsurface mechanisms of active region formation.

\section{Conclusions}\label{sect:conc}
We have identified two distinct samples of emerging active regions: one sample with a persistent magnetic bipole more than one day prior to emergence, and one sample with no distinct magnetic signature until the emergence time. There are 42 active regions in each sample, which may constitute the extremes of a continuum.

We found that both samples of active regions are associated with converging flows prior to emergence, but that the magnitude of the convergence was statistically significantly different, with the sample of active regions with a persistent pre-emergence bipole emerging into strongly converging flows.
We also found a statistical dependence on the post-emergence flux of the active regions, where the sample of active regions with a persistent pre-emergence bipole evolved into lower flux active regions, with an average flux of $(19\pm1.3)~\times 10^{20}$~Mx, and those without evolved into  stronger active regions, with an average flux of $(24\pm1.8) \times 10^{20}$~Mx. 

Furthermore, we found that the higher flux active regions are associated with a diverging flow at the time of emergence, whereas the lower flux active regions did not show any significant flow signature. The ensemble averages of the flows for both samples show the same profile of the diverging flows in time,  offset by about $3 \times 10^{-6}$\ps (and about 30~\ms in the averaged east-west flows).

We have presented a new observational constraint for flux emergence models, and a potential quantity to predict the maximum magnetic flux of an emerging active region. A full interpretation of these intriguing results requires both a more in-depth observational analysis of a broader sample of active regions coupled with numerical simulations of rising flux tubes with a range of magnetic fluxes. This will help to understand whether the flux dependence of the flows in these samples we have identified are distinct or are the extremes of a continuum of EARs governed by a single emergence mechanism.

\begin{acknowledgement}
We acknowledge the Awabakal people, the traditional custodians of the unceded land on which this research was undertaken. Observations courtesy of NASA/SDO and the HMI science teams.
\end{acknowledgement}

\paragraph{Funding Statement}
The HEARs data were originally processed at the German Data Center for SDO, funded by the German Aerospace Center under grant DLR 50OL1701.  
HS is the recipient of an Australian Research Council Future Fellowship Award (project number FT220100330) and this research is partially funded by this grant from the Australian Government.

\paragraph{Competing Interests}

None.

\paragraph{Data Availability Statement}

The HEARs data can be reproduced following the description in this paper and more fully from \cite{Schunkeretal2016}. The results presented in this paper can be fully reproduced following Section~3 and Section~4. Digital data is available through private communication with the authors.

%\endnote in some journals will behave like \footnote; and \printendnotes will not output anything. 
%\printendnotes
%\bibliography{manuscript_final}
\printbibliography

\appendix

\onecolumn
\section{NOAA active region numbers with associated pre-emergence bipoles}\label{app:lbp}

As described in \cite{Schunkeretal2016}, we assigned a number, $P$, to indicate the amount of pre-emergence flux (see Table~\ref{tab:lbp} and \ref{tab:nlbp} in \ref{app:lbp} and \ref{app:nlbp}) based on a visual inspection of the mapped magnetograms. 
A $P$-factor of 0 represents an emergence into a very quiet region; a $P$ factor of 1 or 2 indicates emergence into increasing amounts of magnetic field nearby (but not directly at) the subsequent emergence location; and a $P$-factor of 3 or higher indicates the region may be compromised by pre-existing field at the emergence time and location. These were all evaluated by inspection of the line-of-sight magnetic field maps, and does not appear to correlate with the existence of pre-emergence bipoles.

% TABLE GOES HERE
\begin{longtable}{ l c c c c c | c c | c c }
\caption{Emerging active regions with associated pre-emergence bipoles and their associated control region tracking locations and emergence time \citep[adapted from Table A.1. in each of][]{Schunkeretal2016,Schunkeretal2019}. The left panel of the table lists the NOAA active region number,   emergence time, Carrington latitude, Carrington longitude, central meridian distance (CMD) at the time of emergence and the $P$-factor. The middle panel lists the emergence time and  Carrington longitude of the control region. The right panel lists the difference in $B$-angle, $\Delta B = B0\mathrm{(CR)} - B0\mathrm{(EAR)}$, and the rounded difference in days $\Delta T=t_0(\mathrm{CR}) - t_0(\mathrm{EAR})$. Active regions with a maximum flux larger than the median of the entire HEARS are marked with an asterisk.} \label{tab:lbp} \\
 \, \, AR & emergence time &  lat.  & lon. & CMD  & $P$ & CR emergence time & CR lon.  &  $\Delta B0$ &  $\Delta T$ \\
 \, \, \, \#  & [TAI] & [$^\circ$] & [$^\circ$] & [$^\circ$] &  & [TAI] & [$^\circ$] & [$^\circ$] & [days] \\
\hline
\endfirsthead
\multicolumn{3}{l}{{\bfseries \tablename\ \thetable{} -- continued from previous page}} \\
 \, \,  AR & emergence time &  lat.  & lon. & CMD  & $P$ & CR emergence time & CR lon.  &  $\Delta B0$ &  $\Delta T$ \\
 \, \, \, \#  &[TAI] & [$^\circ$] & [$^\circ$] & [$^\circ$] &  & [TAI] & [$^\circ$] & [$^\circ$] & [days] \\
\hline
\endhead
11066   &  2010.05.02\_23:48:00   &    -26.6   &    208.2   &    -16.8   &  0   &  2010.05.10\_23:48:00   &    102.4   &      0.8   &    7 \\ 
11072*   &  2010.05.20\_17:12:00   &    -15.1   &    314.4   &    -36.1   &  0   &  2010.05.22\_17:12:45   &    288.0   &      0.2   &    2 \\ 
11074   &  2010.05.29\_01:36:00   &     18.6   &    285.4   &     45.3   &  1   &  2010.05.31\_01:36:00   &    258.9   &      0.2   &    1 \\ 
11075   &  2010.05.28\_13:48:00   &    -20.2   &    229.4   &    -17.2   &  1   &  2010.06.11\_13:48:00   &    123.5   &      1.7   &   14 \\ 
11098   &  2010.08.10\_23:12:00   &     13.9   &    300.9   &    -41.2   &  3   &  2010.08.08\_23:12:00   &    327.4   &     -0.1   &   -1 \\ 
11103   &  2010.09.01\_10:12:00   &     26.2   &     85.4   &     26.7   &  4   &  2010.09.10\_00:00:00   &    111.8   &      0.1   &    8 \\ 
11136   &  2010.12.24\_08:24:00   &    -21.4   &     30.4   &     34.3   &  0   &  2011.01.01\_08:23:15   &     56.7   &     -0.9   &    7 \\ 
11141*   &  2010.12.30\_22:36:00   &     34.5   &    267.9   &     -1.4   &  2   &  2010.12.28\_22:36:00   &    294.3   &      0.3   &   -2 \\ 
11143   &  2011.01.06\_01:12:00   &    -22.1   &    145.6   &    -43.3   &  3   &  2011.01.08\_01:12:45   &    119.2   &     -0.2   &    2 \\ 
11148   &  2011.01.17\_02:24:00   &    -27.7   &     65.2   &     21.8   &  0   &  2011.01.22\_02:24:45   &    359.3   &     -0.4   &    5 \\ 
11158*   &  2011.02.11\_01:24:00   &    -19.3   &     35.9   &    -38.8   &  1   &  2011.02.13\_01:23:15   &      9.6   &     -0.1   &    1 \\ 
11182   &  2011.03.27\_04:12:00   &     13.2   &    201.5   &    -12.0   &  4   &  2011.03.25\_04:12:00   &    227.8   &     -0.1   &   -1 \\ 
11198   &  2011.04.21\_14:00:00   &    -25.9   &    272.1   &     33.9   &  1   &  2011.04.23\_14:00:45   &    245.7   &      0.2   &    2 \\ 
11318*   &  2011.10.11\_20:12:00   &     20.9   &     94.9   &    -12.6   &  2   &  2011.10.19\_00:00:00   &     68.5   &     -0.5   &    7 \\ 
11385   &  2011.12.22\_04:12:00   &    -30.5   &    225.3   &    -21.9   &  2   &  2011.11.29\_04:12:00   &     80.4   &      3.0   &  -22 \\ 
11414   &  2012.02.04\_09:24:00   &     -5.4   &     35.7   &     10.8   &  0   &  2012.02.06\_09:23:15   &      9.4   &     -0.1   &    1 \\ 
11446   &  2012.03.22\_17:24:00   &     24.5   &    103.3   &    -18.1   &  0   &  2012.03.14\_17:23:15   &    208.8   &     -0.2   &   -8 \\ 
11510   &  2012.06.18\_20:36:00   &    -16.2   &     17.8   &    -18.8   &  2   &  2012.06.20\_06:21:00   &    271.9   &      0.1   &    1 \\ 
11531*   &  2012.07.25\_11:12:00   &     14.4   &    308.4   &     36.3   &  2   &  2012.07.30\_11:12:00   &    242.3   &      0.4   &    4 \\ 
11547   &  2012.08.16\_09:36:00   &      5.4   &    297.4   &    -44.7   &  3   &  2012.08.18\_09:36:00   &    270.9   &      0.1   &    1 \\ 
11549   &  2012.08.18\_14:12:00   &    -17.8   &    324.1   &     11.0   &  1   &  2012.08.12\_06:21:00   &    350.5   &     -0.3   &   -6 \\ 
11626   &  2012.12.03\_01:36:00   &     12.5   &    299.0   &    -49.2   &  3   &  2012.12.05\_01:36:00   &    272.6   &     -0.3   &    2.0 \\ 
11640*   &  2012.12.29\_15:24:00   &     27.8   &    319.3   &    -38.9   &  0   &  2012.12.31\_15:24:00   &    292.9   &     -0.2   &    2.0 \\ 
11675*   &  2013.02.16\_06:36:00   &     12.5   &     34.2   &    -43.5   &  0   &  2013.02.18\_06:36:45   &      7.9   &     -0.1   &    2.0 \\ 
11702*   &  2013.03.21\_02:12:00   &      8.3   &     14.9   &      9.5   &  0   &  2013.03.23\_02:12:00   &    348.5   &      0.1   &    2.0 \\ 
11750*   &  2013.05.15\_01:48:00   &    -10.3   &    359.8   &      0.5   &  3   &  2013.05.24\_01:48:45   &    240.7   &      1.0   &    9.0 \\ 
11776*   &  2013.06.18\_12:24:00   &     11.7   &    252.1   &    -11.5   &  1   &  2013.06.16\_12:24:45   &    278.5   &     -0.2   &   -2.0 \\ 
11784*   &  2013.07.01\_11:24:00   &    -14.8   &     52.7   &    -39.3   &  3   &  2013.07.03\_11:24:45   &     26.2   &      0.2   &    2.0 \\ 
11813*   &  2013.08.06\_20:00:00   &    -13.1   &    320.7   &    -10.2   &  0   &  2013.08.11\_01:20:15   &    264.9   &      0.3   &    4.2 \\ 
11821   &  2013.08.14\_06:24:00   &      1.3   &    245.4   &     12.7   &  1   &  2013.08.10\_17:20:15   &    292.2   &     -0.2   &   -3.5 \\ 
11829*   &  2013.08.20\_17:00:00   &      4.2   &    190.0   &     42.4   &  3   &  2013.08.23\_17:00:00   &    150.3   &      0.1   &    3.0 \\ 
11831*   &  2013.08.21\_06:48:00   &     13.5   &    165.2   &     25.2   &  2   &  2013.08.24\_06:47:15   &    125.5   &      0.1   &    3.0 \\ 
11833   &  2013.08.22\_08:48:00   &     19.8   &     96.9   &    -28.7   &  4   &  2013.08.26\_12:00:00   &     42.3   &      0.1   &    4.1 \\ 
11867*   &  2013.10.09\_05:00:00   &     23.2   &    180.3   &    -33.7   &  0   &  2013.10.25\_05:00:45   &    329.2   &     -1.2   &   16.0 \\ 
11878   &  2013.10.19\_15:24:00   &     -9.9   &    110.1   &     33.7   &  3   &  2013.10.24\_12:00:00   &     46.0   &     -0.4   &    4.9 \\ 
11915*   &  2013.12.03\_05:48:00   &    -29.6   &    206.9   &     -1.5   &  2   &  2013.11.26\_00:00:00   &    302.3   &      0.9   &   -7.2 \\ 
11946*   &  2014.01.04\_10:36:00   &      9.8   &     99.9   &    -44.3   &  3   &  2013.12.26\_17:20:15   &    214.7   &      1.0   &   -8.7 \\ 
11962   &  2014.01.19\_07:48:00   &    -37.2   &    279.6   &    -28.6   &  0   &  2014.01.21\_07:47:15   &    253.3   &     -0.2   &    2.0 \\ 
11969*   &  2014.01.30\_19:24:00   &    -10.5   &    159.8   &      2.8   &  1   &  2014.01.17\_12:00:00   &    335.1   &      1.1   &  -13.3 \\ 
11992   &  2014.02.25\_20:36:00   &    -20.2   &    137.1   &    -36.8   &  3   &  2014.02.23\_20:35:15   &    163.5   &      0.0   &   -2.0 \\ 
12039   &  2014.04.15\_15:12:00   &     23.9   &    234.8   &    -16.0   &  1   &  2014.04.18\_15:12:45   &    195.2   &      0.2   &    3.0 \\ 
12105   &  2014.06.28\_23:24:00   &     -7.1   &    307.8   &    -39.7   &  2   &  2014.06.26\_23:24:00   &    334.3   &     -0.2   &   -2.0 \\ 
\end{longtable}

%AR_Numbers_LBP.txt created by C. Alley from visual inspection of the line of sight magnetograms averaged over six hours
%\tablefoot{From the left column, the table lists the NOAA active region number,   emergence time, Carrington latitude, Carrington longitude, central meridian distance (CMD) at the time of emergence and the $P$-factor. Columns 7 and 8 list the emergence time and  Carrington longitude of the control region. The last two columns lists the difference in $B$-angle, $\Delta B = B0\mathrm{(CR)} - B0\mathrm{(EAR)}$, and the rounded difference in days $\Delta T=t_0(\mathrm{CR}) - t_0(\mathrm{EAR})$.\\ \tablefoottext{*}{Active regions with a maximum flux larger than the median of the entire HEARS.}}

\section{NOAA active region numbers without associated pre-emergence bipoles}\label{app:nlbp}

% TABLE GOES HERE
\begin{longtable}{ l c c c c c | c c | c c }
\caption{Emerging active region and control region tracking locations and emergence time \citep[adapted from Table A.1. in each of][]{Schunkeretal2016,Schunkeretal2019}. The left panel of the table lists the NOAA active region number,   emergence time, Carrington latitude, Carrington longitude, central meridian distance (CMD) at the time of emergence and the $P$-factor. The middle panel lists the emergence time and  Carrington longitude of the control region. The right panel lists the difference in $B$-angle, $\Delta B = B0\mathrm{(CR)} - B0\mathrm{(EAR)}$, and the rounded difference in days $\Delta T=t_0(\mathrm{CR}) - t_0(\mathrm{EAR})$. Active regions with a maximum flux larger than the median of the entire HEARS are marked with an asterisk.} \label{tab:nlbp} \\
 \, \, AR & emergence time &  lat.  & lon. & CMD  & $P$ & CR emergence time & CR lon.  &  $\Delta B0$ &  $\Delta T$ \\
 \, \, \, \#  & [TAI] & [$^\circ$] & [$^\circ$] & [$^\circ$] &  & [TAI] & [$^\circ$] & [$^\circ$] & [days] \\
\hline
\endfirsthead
\multicolumn{3}{l}{{\bfseries \tablename\ \thetable{} -- continued from previous page}} \\
 \, \,  AR & emergence time &  lat.  & lon. & CMD  & $P$ & CR emergence time & CR lon.  &  $\Delta B0$ &  $\Delta T$ \\
 \, \, \, \#  &[TAI] & [$^\circ$] & [$^\circ$] & [$^\circ$] &  & [TAI] & [$^\circ$] & [$^\circ$] & [days] \\
\hline
\endhead
11070   &  2010.05.05\_03:24:00   &     20.7   &    195.0   &     -1.5   &  1   &  2010.05.09\_00:00:00   &     89.3   &      0.4   &    3 \\ 
11079*   &  2010.06.08\_08:24:00   &    -26.0   &    118.5   &     14.5   &  1   &  2010.06.10\_08:23:15   &     92.1   &      0.2   &    1 \\ 
11080*   &  2010.06.10\_02:12:00   &    -23.1   &    109.2   &     28.3   &  2   &  2010.06.12\_02:12:00   &     82.8   &      0.2   &    2 \\ 
11081*   &  2010.06.11\_07:12:00   &     24.0   &    100.5   &     35.6   &  1   &  2010.06.13\_07:12:00   &     74.0   &      0.2   &    1 \\ 
11086   &  2010.07.04\_08:36:00   &     17.8   &    152.0   &     32.2   &  1   &  2010.07.06\_08:36:00   &    125.5   &      0.2   &    1 \\ 
11122   &  2010.11.06\_01:12:00   &     13.8   &    261.3   &    -11.3   &  0   &  2010.11.08\_01:12:45   &    235.0   &     -0.2   &    2 \\ 
11174*   &  2011.03.16\_20:12:00   &     21.3   &     10.7   &     21.0   &  2   &  2011.03.14\_20:12:00   &     37.0   &     -0.0   &   -2 \\ 
11194   &  2011.04.13\_05:12:00   &    -31.8   &      8.9   &     20.3   &  3   &  2011.04.15\_05:11:15   &    342.5   &      0.1   &    1 \\ 
11199*   &  2011.04.25\_18:36:00   &     21.2   &    187.3   &      4.5   &  2   &  2011.05.09\_18:36:00   &      2.3   &      1.4   &   14 \\ 
11209   &  2011.05.08\_04:48:00   &     34.8   &    358.9   &    -19.6   &  1   &  2011.05.10\_04:48:00   &    332.5   &      0.2   &    1 \\ 
11211   &  2011.05.08\_15:24:00   &    -13.6   &     16.2   &      3.4   &  1   &  2011.05.03\_15:24:00   &     82.3   &     -0.5   &   -4 \\ 
11273   &  2011.08.16\_13:24:00   &    -17.1   &    111.0   &    -19.8   &  2   &  2011.09.08\_13:24:00   &     44.9   &      0.6   &   22 \\ 
11297*   &  2011.09.13\_17:48:00   &    -17.6   &    152.3   &     33.9   &  1   &  2011.09.08\_17:48:45   &    218.3   &      0.0   &   -4 \\ 
11300*   &  2011.09.17\_03:48:00   &     24.2   &     92.3   &     19.0   &  0   &  2011.09.24\_00:00:00   &     65.9   &     -0.2   &    6 \\ 
11311*   &  2011.10.03\_16:36:00   &    -12.8   &    177.2   &    -37.9   &  0   &  2011.10.23\_16:36:00   &    273.3   &     -1.4   &   20 \\ 
11322*   &  2011.10.15\_14:24:00   &    -27.0   &    103.5   &     45.5   &  1   &  2011.10.01\_14:24:00   &     37.5   &      0.9   &  -13 \\ 
11331*   &  2011.10.22\_18:36:00   &     10.1   &      5.6   &     42.3   &  1   &  2011.10.20\_18:36:00   &     32.0   &      0.2   &   -2 \\ 
11334*   &  2011.10.30\_00:36:00   &     11.3   &    187.9   &    -39.8   &  2   &  2011.10.28\_00:36:00   &    214.3   &      0.2   &   -2 \\ 
11397   &  2012.01.12\_22:36:00   &    -20.5   &    277.1   &    -43.3   &  1   &  2012.01.30\_22:36:45   &     92.8   &     -1.6   &   18 \\ 
11416*   &  2012.02.08\_18:24:00   &    -18.5   &    287.6   &    -39.8   &  1   &  2012.02.16\_18:23:15   &    182.2   &     -0.4   &    7 \\ 
11431*   &  2012.03.04\_13:12:00   &    -28.7   &     16.3   &     15.4   &  1   &  2012.03.09\_13:11:15   &    310.5   &      0.0   &    4 \\ 
11437   &  2012.03.16\_16:12:00   &    -34.3   &    167.7   &    -33.4   &  1   &  2012.03.14\_16:12:45   &    194.1   &     -0.0   &   -1 \\ 
11560*   &  2012.08.29\_11:36:00   &      2.9   &    125.4   &    -43.8   &  1   &  2012.08.21\_11:35:15   &    231.1   &     -0.2   &   -8 \\ 
11561   &  2012.08.30\_01:48:00   &    -12.4   &    132.5   &    -28.9   &  1   &  2012.09.10\_01:48:45   &    347.2   &      0.1   &   11 \\ 
11570   &  2012.09.11\_19:00:00   &    -12.8   &     10.4   &     16.9   &  0   &  2012.09.13\_18:59:15   &    344.0   &     -0.0   &    1 \\ 
11624   &  2012.11.27\_12:12:00   &     20.7   &     32.5   &    -29.0   &  1   &  2012.11.23\_00:00:00   &    247.5   &      0.6   &   -4 \\ 
11645*   &  2013.01.02\_20:12:00   &    -13.3   &    290.4   &    -12.4   &  0   &  2012.12.29\_20:12:00   &    343.1   &      0.5   &   -4.0 \\ 
11696*   &  2013.03.11\_10:24:00   &      4.4   &     90.5   &    317.8   &  1   &  2013.03.20\_12:00:00   &    331.0   &      0.2   &    9.1 \\ 
11697   &  2013.03.13\_13:00:00   &     14.7   &    107.7   &      2.8   &  1   &  2013.03.22\_12:00:00   &    349.6   &      0.2   &    9.0 \\ 
11699*   &  2013.03.17\_00:24:00   &    -15.8   &     91.4   &     32.3   &  0   &  2013.03.05\_12:00:00   &    243.2   &     -0.1   &  -11.5 \\ 
11706   &  2013.03.27\_01:24:00   &     -6.5   &    268.7   &    -18.0   &  1   &  2013.04.03\_01:23:15   &    176.4   &      0.4   &    7.0 \\ 
11707   &  2013.03.28\_11:48:00   &    -10.7   &    229.0   &    -38.8   &  0   &  2013.03.26\_11:48:00   &    255.4   &     -0.1   &   -2.0 \\ 
11718*   &  2013.04.05\_15:24:00   &     22.0   &    109.6   &    -50.6   &  0   &  2013.04.03\_15:24:00   &    136.0   &     -0.1   &   -2.0 \\ 
11786   &  2013.07.02\_00:00:00   &    -32.1   &     53.7   &    -31.4   &  0   &  2013.07.04\_00:00:00   &     27.2   &      0.2   &    2.0 \\ 
11824*   &  2013.08.17\_07:36:00   &    -14.8   &    194.8   &      2.4   &  1   &  2013.08.26\_12:00:00   &     73.4   &      0.3   &    9.2 \\ 
11849*   &  2013.09.19\_13:00:00   &     20.9   &     75.3   &    -38.2   &  1   &  2013.09.16\_12:00:00   &    115.5   &      0.1   &   -3.0 \\ 
11910*   &  2013.11.27\_13:12:00   &      1.5   &    276.3   &     -7.1   &  1   &  2013.11.25\_13:11:15   &    302.7   &      0.3   &   -2.0 \\ 
11978*   &  2014.02.10\_07:24:00   &      5.6   &     34.0   &     15.3   &  1   &  2014.01.31\_07:24:00   &    165.7   &      0.6   &  -10.0 \\ 
12041   &  2014.04.15\_15:36:00   &    -20.7   &    262.3   &     11.7   &  0   &  2014.04.13\_12:00:00   &    290.7   &     -0.2   &   -2.2 \\ 
12078   &  2014.05.31\_00:48:00   &    -18.4   &    327.4   &    -43.2   &  1   &  2014.05.28\_12:00:00   &      1.0   &     -0.3   &   -2.5 \\ 
12118   &  2014.07.17\_17:24:00   &      7.0   &    113.3   &     13.9   &  0   &  2014.07.16\_12:00:00   &    129.5   &     -0.1   &   -1.2 \\ 
12119*   &  2014.07.18\_11:12:00   &    -22.1   &     66.8   &    -22.8   &  1   &  2014.07.22\_11:12:00   &     13.9   &      0.4   &    4.0 \\ 
\end{longtable}

%AR_Numbers_nLBP.txt created by C. Alley from visual inspection of the line of sight magnetograms averaged over six hours; double checked 
%\tablefoot{From the left column, the table lists the NOAA active region number,   emergence time, Carrington latitude, Carrington longitude, central meridian distance (CMD) at the time of emergence and the $P$-factor. Columns 7 and 8 list the emergence time and  Carrington longitude of the control region. The last two columns list the difference in $B$-angle, $\Delta B = B0\mathrm{(CR)} - B0\mathrm{(EAR)}$, and the rounded difference in days $\Delta T=t_0(\mathrm{CR}) - t_0(\mathrm{EAR})$. \\\tablefoottext{*}{Active regions with a maximum flux larger than the median of the entire HEARS.}}

\twocolumn

\section{Averaged flow maps of control regions for both samples}\label{app:aveqs}
\begin{figure*}
	\includegraphics[width=0.9\textwidth]{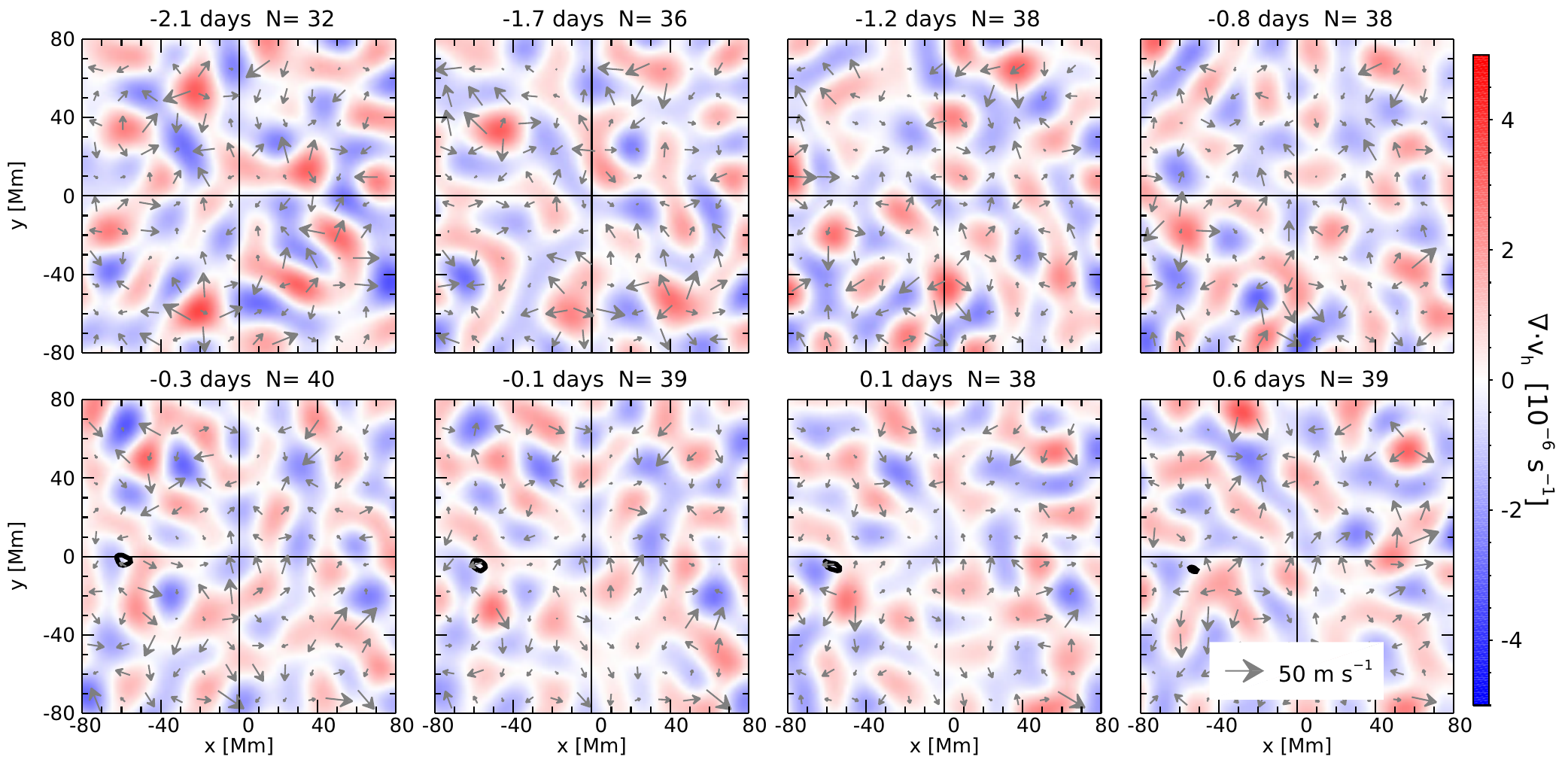}
    \caption{Averaged divergence flow maps of the control regions associated with pre-emergence bipole regions.  Blue represents converging flows and red represents diverging flows. The arrows indicate the direction and magnitude of the flows. Solid (dashed) black lines contour  the $-20$~G ($+20$~G) of the averaged line-of-sight magnetic field maps. There are no significant flows. }
    \label{fig:avelbpqsmaps}
\end{figure*}
% ave_maps_EARS_lists.pro
% image_ave_EARS_lists.pro
% figure_ave_EAR_lists.pro
\begin{figure*}
	\includegraphics[width=0.9\textwidth]{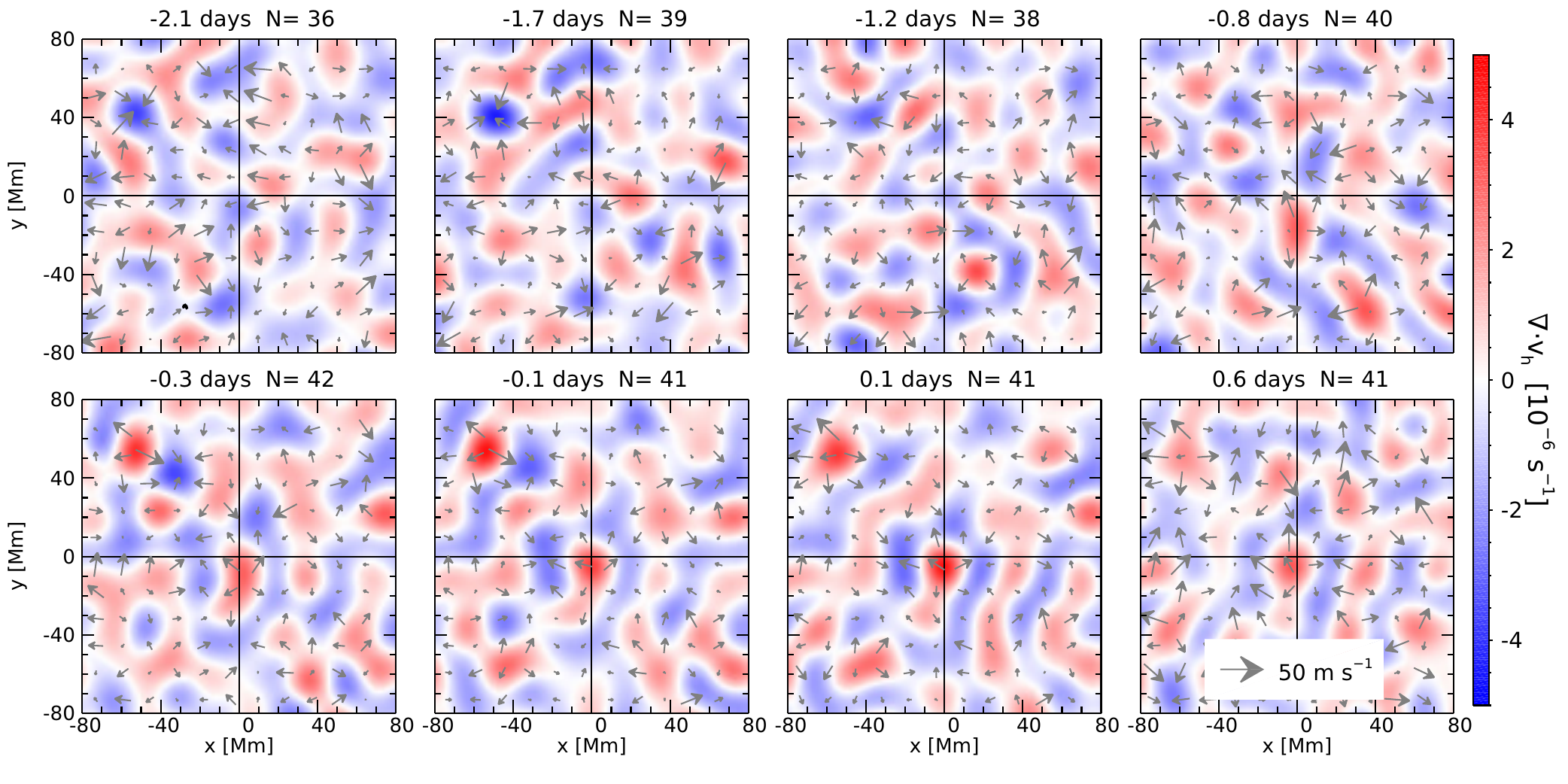}
    \caption{Averaged divergence flow maps of the control regions associated with active regions without pre-emergence bipoles. Blue represents converging flows and red represents diverging flows. The arrows indicate the direction and magnitude of the flows.  There is a diverging flow at the centre of the map near the artificial emergence time, however, neither the size nor magnitude is significantly different than other surrounding regions. We note that most of the Sun's surface consists of supergranulation cells of diverging flows, and so this is statistically not unexpected. }
    \label{fig:avenlbpqsmaps}
\end{figure*}
% ave_maps_EARS_lists.pro
% image_ave_EARS_lists.pro
% figure_ave_EAR_lists.pro
%%%%%%%%%%%%%%%%%%%%%%%%%%%%%%%%%%%%%%%%%%%%%%%%%%

\end{document}